\documentclass{article}
\usepackage{amsmath}
\usepackage{amsfonts}
\begin{document}

\title{Landau problem with a general time-dependent electric field}

\author{J. Chee\\\\Department of Physics, School of Science, \\Tianjin Polytechnic University,
Tianjin 300387, China}

 \date{July, 2012}
 \maketitle
 \begin{abstract}
 The time evolution is studied for the Landau level problem with a general time dependent
 electric field ${\bf E}(t)$ in a plane perpendicular to the magnetic field.
 A general and explicit factorization of the
 time evolution operator is obtained with each factor having a clear physical interpretation.
 The factorization consists of a geometric factor (path-ordered magnetic translation),
 a dynamical factor generated by the usual time-independent Landau Hamiltonian, and a
 nonadiabatic factor that determines the transition probabilities among
 the Landau levels. Since the path-ordered magnetic translation and the nonadiabatic factor are,
 up to completely determined numerical phase factors, just
 ordinary exponentials whose exponents are explicitly expressible in terms of the canonical variables, all of the
 factors in the factorization are explicitly constructed. The
 numerical phase factors are quantum mechanical in nature and could be of significance in interference experiments.
 The factorization is unique from the point of
 view of the quantum adiabatic theorem and provides a demonstration of
 how the quantum adiabatic theorem (incorporating the Berry phase phenomenon) is realized when infinitely
 degenerate energy levels are involved. Since the factorization separates
 the effect caused by the electric field into a geometric factor and a nonadiabatic factor,
 it makes possible to calculate the nonadiabatic transition probabilities near the
 adiabatic limit. A formula for matrix elements that
 determines the mixing of the Landau levels for a general, non-adiabatic evolution is also provided by the factorization.
 \end{abstract}

\pagebreak

\section{Introduction}
The magnetic translation concept is much discussed in the literature
\cite{brown1,zak1,zak2,tanimura,chee,florek,exner}. The basic
example is the quantum mechanics of a charged particle moving in a
two-dimensional plane perpendicular to a magnetic field, i.e., the
Landau level problem. The Hamiltonian does not commute with the
usual translation operator generated by the canonical momentum. The
translation symmetry of the physical situation is realized through
magnetic translation. However, in this simple context, the solution
of the problem (energy levels, etc.) does not seem to rely on this
concept in any essential way. The magnetic translation concept has
been used \cite{zak4,brown2,kohmoto,wiegmann} in condensed matter
physics including Bloch electrons in a magnetic field where there is
a lattice potential present. Yet we believe the essential role of
magnetic translation in exploring time-dependent problems have yet
to be fully explored.

In this paper, we study the quantum mechanical problem of a charged
particle moving in a two-dimensional plane subjected to a uniform
magnetic field perpendicular to the plane and a spatially uniform
but time-dependent in-plane electric field ${\bf E}(t)$. The
Hamiltonian for such a system is the sum of the usual Landau level
Hamiltonian $H_0$ and the potential energy of the charge in the
electric field:
\begin{eqnarray}
H&=&H_0+V({\bf x}, t),\nonumber\\&=&\frac{1}{2m}\big[{\bf
p}-\frac{q}{c}{\bf A}({\bf x})\big]^2 -qE_1(t)x_1-qE_2(t)x_2,
\end{eqnarray}
where ${\bf A}({\bf x})$ satisfies $\nabla\times{\bf A}({\bf
x})=B{\bf e}_3$. The situation contains the usual Landau level
problem as a special case, where $E_1(t)=E_2(t)=0$. It also contains
the special case for which the uniform electric field is constant in
time, and the case \cite{budd} for which ${\bf E}(t)$ depends on
time but is along a fixed direction. These cases are all known in
the literature. Our purpose here is to study the general case with
${\bf E}(t)=E_1(t){\bf e}_1+E_2(t){\bf e}_2$ being a general
time-dependent electric field. This general situation cannot be
reduced to the sum of two special cases and it will be shown that it
has new features of its own. In particular, we will show that a
path-ordered magnetic translation, which is an element of the
magnetic translation group, plays an essential role in describing
the evolution of the system. This path-ordered magnetic translation
contains a numerical phase factor that is nontrivial only when the
electric field changes direction with time. Our main result is a
factorization of the time-evolution operator into three factors,
each having a clear physical meaning. The method and result seem to
be quite natural that it is possible that they may find applications
in studying more complicated situations such as the Landau problem
on a cylinder \cite{date} or a torus, or when there is a lattice
potential present.

First of all, from the physical picture that underlies the Hall
effect, the circular orbit of a charged particle in mutually
perpendicular electric and magnetic fields exhibits motion in the
direction of ${\bf E}\times{\bf B}$. It is easy to conjecture that,
in the case where ${\bf E}(t)$ is a general time-dependent electric
field, the instantaneous velocity of the global motion of the
circular orbit is $cE(t)/B$, in the direction of ${\bf E}\times{\bf
B}$. Therefore, the position of the center of the circular orbit at
time $t$ is described by a parameter
\begin{equation}
{\bf R}(t)=\frac{c}{B}\int_{0}^{t}\big(E_2(s){\bf e}_1-E_1(s){\bf
e}_2\big)ds,
\end{equation}
where we have taken the initial position ${\bf R}(0)$ to be at the
origin of the parameter space: ${\bf R}(0)={\bf 0}$.

In this paper, we aim to find a general factorization of the
time-evolution operator $U(t, 0)$ corresponding to $H$ in terms of
three factors: a geometrical factor describing the physical
displacement of a quantum wave by the amount of ${\bf R}(t)$, a
dynamical factor generated by the usual time-independent Landau
Hamiltonian $H_0$, and another factor that describes the mixing of
the Landau levels of $H_0$. The factorization is general in the
sense that it is valid for a general time variation of the function
${\bf R}(t)$ and therefore of ${\bf E}(t)$.

We comment here that previous work on the special cases, such as the
one in \cite{budd} has not relied on the magnetic translation
concept. This is due to the fact that if the electric field is along
a fixed direction, say in the $x_1$ direction, one chooses the
Landau gauge where ${\bf A}({\bf x})=(0, Bx,0)$ and furthermore
specify the constant of motion $p_2$ to be $0$ and the problem is
then simplified. Such a simplification also implies that the quantum
states under consideration have $p_2=0$. For such a restriction, the
geometric operator found in the present paper is equal to the
identity. So the only effects caused by the electric field are
nonadiabatic transitions. No such simplification exists if the
electric field changes direction with time. In order to treat the
general case, the magnetic translation concept is necessary. In
fact, even in the special case, if one does not make the special
restrictions, there should be a magnetic translation along the fixed
direction of $x_2$. We also note that upon choosing a specific
gauge, the Hamiltonian considered here can be cast in a quadratic
form. For such type of time-dependent Hamiltonians, there is a
method that uses the corresponding classical solutions to construct
the propagator of the quantum problem \cite{combescure,dodonov}. While such
an approach could be useful as a general theory, it does not seem to
point to a factorization of the time evolution operator which, in
specific contexts, can make the time evolution transparent.

\section{Factorization of the time-evolution operator}
In this section, we first give a discussion on the properties of the
variables $(\pi_1, \pi_2)$ and $(w_1, w_2)$ which can be expressed
in terms of the canonical variables $\bf x$ and $\bf p$. Then we use
these variables to construct a factorization of the time evolution
operator. Although the commutation relations among these variables
hold in any picture, we assume throughout this paper that when these
variables appear in the exponents, they are Schr{\"{o}}dinger
picture variables.

In the usual Landau problem $H_0$, the kinematical momentum is
$\pi_\mu=p_\mu-\frac{q}{c}A_\mu({\bf x}),\ \mu=1, 2$, where
$A_\mu({\bf x})$ is the vector potential in arbitrary gauge, though
for simplicity we assume that $A_\mu({\bf x})$ does not depend on
time. Define
\begin{equation}
w_{1}=\pi_1-\frac{qB}{c}x_2=-\frac{qB}{c}c_2, \ \ \ \
w_{2}=\pi_2+\frac{qB}{c}x_1=\frac{qB}{c}c_1.
\end{equation}
In classical mechanics, $(c_1, c_2)$ is the center of the circular
motion of the charged particle in the magnetic field. In quantum
mechanics, because of the canonical commutation relations, $c_1$ and
$c_2$ do not commute. We have the following commutation relations
\begin{equation}
[\pi_1, \pi_2]=i\hbar qB/c,\ \ \  [w_{1}, w_{2}]= -i\hbar qB/c,\ \ \
[\pi_\mu, w_{\nu}]=0.
\end{equation}
To realize a translation ${\bf x}\rightarrow {\bf x}+{\bf R}(t)$,
where ${\bf R}(t)$ traverses through a path $C_{\bf R}$ in parameter
space, one may use the ordinary translation operator $\exp(-ip_\mu
R_{\mu}(t)/\hbar)$. (Summation over repeated indices is assumed.)
However, the ordinary translation does not commute with $\pi_\mu$.
The magnetic translation operator, defined as
\begin{equation}
\overline{M}({\bf R})=\exp(-iw_{\mu} R_{\mu}/\hbar),
\end{equation}
is a generalization of the ordinary translation operator when a
magnetic field is present: It physically translates a quantum wave
because it commutes with the kinematical momentum, and in the
simplest Landau system $H_0$, it preserves the energy. Note that
unlike ordinary translation, a distinction between $\exp(-iw_{\mu}
R_{\mu}/\hbar)$ and the path-ordered magnetic translation
\begin{equation}
M(C_{\bf R})=P\exp(-i{\hbar}^{-1}w_{\mu} \int_{C_{\bf
R}}dR_{\mu})=P\exp(-i{\hbar}^{-1}w_{\mu} R_{\mu})
\end{equation}
has to be made, for a general path $C_{\bf R}$ that is not in a
straight line, because $w_1$ and $w_2$ do not commute. (We assume
$R_{\mu}(0)=0$.) Because of the simple commutation relation $[w_{1},
w_{2}]= -i\hbar qB/c$, their difference is a numerical phase factor,
i.e.
\begin{equation}
M(C_{\bf R})=e^{i\beta(C_{\bf R})}\exp(-iw_{\mu} R_{\mu}/\hbar),
\end{equation}
where $\beta(C_{\bf R})$ is a real number determined by the path
$C_{\bf R}$ traversed by ${\bf R}(t)$. In particular, for a closed
path, $\beta(C_{\bf R})$ is equal to $-\frac{q\phi}{\hbar c}$, where
$\phi$ is the magnetic flux enclosed by the loop $C$. This follows
from the definition of a path-ordered exponential, the commutation
relation $[w_{1}, w_{2}]= -i\hbar qB/c$ and the formula
$e^Ae^B=e^{A+B}e^{\frac{1}{2}[A, B]}$ for $A$ and $B$ commuting with
$[A, B]$. For an open path ${\bf R}(t)$, the flux is through the
area enclosed by the path and the straight line pointing from the
end point ${\bf R}(t)$ to the initial point ${\bf R}(0)={\bf 0}$. In
general we have
\begin{equation}
\beta(C_{\bf R})=-\frac{qB}{\hbar c}{\bf
e}_3\cdot\frac{1}{2}\int_{C_{\bf R}}{\bf R}\times d{\bf
R}=-\frac{qB}{\hbar c}S.
\end{equation}
Or, if we denote $R(t)=R_1(t)+iR_2(t)$, then
\begin{equation}
\beta(C_{\bf R})=-\frac{qB}{\hbar c}\frac{(-i)}{4}\int_{C_{\bf
R}}(R^{\ast}dR-RdR^\ast).
\end{equation}
Another path-ordered exponential that is relevant to our purpose is
\begin{equation}
J(C_u)=P\exp\big(\pi u/\hbar-\pi^\dagger u^\ast/\hbar\big),\ \ \
(\pi=\pi_1+i\pi_2)
\end{equation}
where the path $C_u$ is the one traversed by $u(t)$ in a complex
$u$-plane. Similar to the path-ordered magnetic translation, this
path-ordered exponential can be evaluated by using the formula
$[\pi, \pi^\dagger]=2\hbar qB/c$. We have
\begin{equation}
J(C_u)=e^{i\gamma(C_u)}\exp\big(\pi u/\hbar-\pi^\dagger
u^\ast/\hbar\big),
\end{equation}
where
\begin{equation}
\gamma(C_u)=i\frac{qB}{\hbar
c}\int_{C_u}(u^{\ast}du-udu^\ast)=-\frac{qB}{\hbar c}4S(C_u).
\end{equation}
$S(C_u)$ is the area enclosed by the path traversed by $u(t)$ in the
complex $u$-plane and the straight line connecting the end and
initial points of the path.

The basic idea that leads to the factorization of $U(t,0)$ is to
switch from the Schr{\"{o}}dinger picture to the Heisenberg picture
first. Furthermore, we observe that the Heisenberg equations of
motion decouple from each other for the variables $\pi_\mu(t)$ and
$w_\mu(t)$, unlike the equations of motion for the canonical
variables. It is the behavior of $\pi_\mu(t)$ and $w_\mu(t)$ instead
of that of $p_\mu(t)$ and $x_\mu(t)$ that leads to the factorization
of $U(t, 0)$; namely we find an operator $O(t, 0)$ in a factorized
form that recovers the evolution of $\pi_\mu(t)$ and $w_\mu(t)$
through $\pi_\mu(t)=O^{\dagger}(t, 0)\pi_\mu O(t, 0)$ and
$w_\mu(t)=O^{\dagger}(t, 0)w_\mu O(t, 0)$. We then verify that this
operator is in fact $U(t,0)$ by a computation that shows it
satisfies the Schr\"{o}dinger equation with the initial condition
$U(0,0)=I$. These steps are presented in detail in Appendix A. The
exact and general factorization of the time evolution operator $U(t,
0)$ is then determined to be
\begin{equation}
U(t, 0)=M(C_{\bf R})D(t)J(C_u),
\end{equation}
where $D(t)=\exp(-iH_0t/\hbar)$ is the evolution generated by the
usual Landau Hamiltonian; $M(C_{\bf R})$ describes path-ordered
magnetic translation of a wave corresponding to the path $C_{\bf R}$
traversed by ${\bf R}(t)=\frac{c}{B}\int_{0}^{t}\big(E_2(s){\bf
e}_1-E_1(s){\bf e}_2\big)ds$. These two operators commute with each
other, so no energy is gained or lost by the action on the
wave-function of a magnetic translation.
The complex parameter $u$ that determines the
operator $J(C_u)$ is given by
\begin{equation}
u(t)=\frac{i}{2}\int_{0}^{t}e^{-i\omega s}\frac
{d}{ds}R^{\ast}(s)ds=\frac{-c}{2B}\int_{0}^{t}e^{-i\omega
s}E^{\ast}(s)ds,
\end{equation}
where $R^{*}(s)=R_1(s)-iR_2(s)$, $E^{*}(s)=E_1(s)-iE_2(s)$, and
$\omega=qB/(mc)$.

We see that the operator $\exp\big(\pi u/\hbar-\pi^\dagger
u^\ast/\hbar\big)$ in $J(C_u)$ mixes different energy levels of
$H_0$. This is due to the fact that $\pi^{\dagger}$ and $\pi$ have
the meaning of being proportional to the creation and annihilation
operators: the Hamiltonian $H_0$ can be written as
$H_0=\hbar\omega(a^{\dagger}a+1/2)$, where $a^{\dagger}=(\hbar
k)^{-1} \pi^{\dagger}, a=(\hbar k)^{-1}\pi$, with $(\hbar
k)^{-1}=\sqrt{c/(2qB\hbar)}$. From this we see that the operator
$J(C_u)$ has the meaning that, when acting on any of the ground
states of $H_0$, it generates a coherent state associated with the
minimization of $\Delta \pi_1\cdot\Delta \pi_2$ rather than the
minimization of $\Delta x_1\Delta p_1$, or of $\Delta x_2\Delta
p_2$. The coherent state nature of $J(C_u)|\Psi_0\rangle$, where
$|\Psi_0\rangle$ is a ground state of $H_0$, is then preserved under
the action of $M(C_{\bf R})D(t)$ in the time-evolution operator.

\section{Gauge invariance and explicitness of the factorization}
Apart from the numerical phase factors $e^{i\beta(C_{\bf R})}$ and
$e^{i\beta(C_u)}$, the operators $M(C_{\bf R})$ and $J(C_u)$ are
generated by $(w_1, w_2)=(-\frac{qB}{c}c_2, \frac{qB}{c}c_1)$ and
$(\pi_1, \pi_2)$ respectively. In the system $H_0$, the physical
meaning of these generators are given by the ``center of the
circular orbit" and the kinematical momentum which are gauge
independent. However, one must be careful in claiming the
factorization gauge invariant in the most general way since the
derivation of the factorization relies on the condition that
$\pi_\mu$ and $w_\mu$ in the exponents are time-independent
Schr\"{o}dinger variables and as such, $A_\mu$ is assumed to depend
on ${\bf x}$ only. For example, in the symmetric gauge where ${\bf
A}({\bf x})=\frac{B}{2}{\bf e}_3\times{\bf x}$, we have
$\pi_{1}=p_1+\frac{qB}{2c}x_2$, $\pi_{2}=p_2-\frac{qB}{2c}x_1$,
$w_{1}=p_1-\frac{qB}{2c}x_2$ and $w_{2}=p_2+\frac{qB}{2c}x_1$.
Therefore, the three factors in the factorization are all explicit
functions of the canonical variables once a specific gauge is
chosen.

Consider the example of a rotating electric field. We have
$R(t)=R_0(e^{-i\nu t}-1)$ or $E(t)=(iB/c)\dot{R}(t)=(\nu
B/c)R_0e^{-i\nu t}=E_0e^{-i\nu t}$. By the formulas (9), (12) and
(14), we have $\beta(C_{\bf R})=(qB/\hbar c)\frac{1}{2}{R_0}^2(\nu
t-\sin\nu t)$, $u(t)=(-R_0\nu/2)(e^{-i\omega t+i\nu
t}-1)/(-i\omega+i\nu)$, and $\gamma(C_u)=(qB/\hbar
c)({R_0}^2/2)(\frac{\nu}{\omega-\nu})^2[(\omega-\nu)t-\sin(\omega-\nu)t]$.
For the case of $\nu\rightarrow\omega$, we have $u(t)=-R_0\nu t/2$,
and $\gamma(C_u)=0$. The expressions for $M(C_{\bf R})$ and $J(C_u)$
are therefore obtained according to (7) and (11). In this particular
example, the result implies that resonance happens at $\nu=\omega$
where $u(t)$ increases linearly with time, thereby causing rapid
transitions among the Landau levels. This example will be further
commented on at the end of the paper.

\section{Physical implications of the factorization}
The general factorization of $U(t, 0)$ demonstrates that a
path-ordered magnetic translation is a natural concept associated
with a charged particle in a time-dependent electric field and a
uniform perpendicular magnetic field. The same can be said of the
path ordered exponential $J(C_u)$ that describes the mixing of the
Landau levels. The factorization also has a natural connection with
the quantum adiabatic theorem as discussed in the next section. The
numerical phase factors $e^{i\gamma(C_u)}$ in $J(C_u)$ and
$e^{i\beta(C_{\bf R})}$ in $M(C_{\bf R})$, which distinguish the
path-ordered exponentials from the direct exponentials, represent
pure quantum effects that have no classical origin and they could be
of consequences in interference experiments. It is obvious that the
factor $e^{i\beta(C_{\bf R})}$ is nontrivial only when the electric
field changes direction with time. This numerical factor contains
the adiabatic Berry phase in the usual sense as discussed in the
next section. One may want to draw an analogy with the Aharonov-Bohm
phase even though the magnetic field interacts directly with the
particle here. However, the phase $\beta(C_{\bf
R})=-\frac{q\phi}{\hbar c}$ that results from a path-ordered
magnetic translation around a closed path $C_{\bf R}$ has an
opposite sign from the Aharonov-Bohm phase. The derivation of the
latter from the point of view of the geometric phase is given in
Berry's paper \cite{berry}.

\section{Relation to the quantum adiabatic theorem}
The factorization can be viewed from a different though equivalent
perspective. The Hamiltonian $H$ can be recast in the form
\begin{equation}
H_L=\frac{1}{2m}\big[{\bf p}-\frac{q}{c}{\bf A}_L(x, {\bf
R})\big]^2,
\end{equation}
through the gauge transformation
\begin{equation}
\Psi_L({\bf x},t)= \exp[-i\frac{q}{\hbar c}\chi({\bf x},{\bf
R})]\Psi({\bf x},t),
\end{equation}
\begin{equation}
{\bf A}_L(x, {\bf R})={\bf A}({\bf x})-{\bf \nabla}\chi({\bf x},{\bf
R})={\bf A}({\bf x})-B{\bf e}_3\times{\bf R}(t),
\end{equation}
where
\begin{equation}
\chi({\bf x},t)= -BR_2(t)x_1+BR_1(t)x_2.
\end{equation}
Because $\Psi_L({\bf x},0)=\Psi({\bf x},0)$, the time-evolution
operators $U_L(t, 0)$ and $U(t, 0)$ are related by
\begin{equation}
U_L(t,0)=\exp[-i\frac{q}{\hbar c}\chi({\bf x},{\bf R})]U(t, 0).
\end{equation}
The Landau level Hamiltonian $H_L(t)$, unlike the gauge equivalent
$H(t)$, has energy eigenvalues $E_n=\hbar\omega(n+1/2)$, where
$\omega=qB/(mc)$, that depend on $B$ only and is independent of the
time variation of $H_L(t)$. Therefore, it is in the gauge of
$H_L(t)$ that $D(t)$ carries the dynamical phase factor of adiabatic
evolutions of eigenstates of the Hamiltonian. This should relate to
the quantum adiabatic theorem, where $G(C_{\bf
R})=\exp[-i\frac{q}{\hbar c}\chi({\bf x},{\bf R})]M(C_{\bf R})$ is a
geometrical operator completely determined by the path of ${\bf
R}(t)$ which brings an initial eigenstate of $H_L(0)$ to an
instantaneous eigenstate of $H_L(t)$, and $J(C_u)$ describes
nonadiabatic transitions. This is an example of the quantum
adiabatic theorem \cite{messiah} where all of the three factors of
the time-evolution operator are explicitly constructed and where the
energy levels of the instantaneous Hamiltonian are infinitely
degenerate.

Recall that the usual quantum adiabatic theorem (and also
generalizations \cite{avron})incorporating the Berry phase
phenomenon is essentially a factorization of the time-evolution
operator into three pieces: a geometric factor that embodies a Berry
phase, a usual dynamical factor, and a nonadiabatic factor that
approaches the identity operator in the adiabatic limit
\cite{messiah}. In our case, once a specific gauge is chosen, the
factors $G(C_{\bf R})=\exp[-i\frac{q}{\hbar c}\chi({\bf x},{\bf
R})]M(C_{\bf R})$ and $J(C_u)$ are exponentials of explicit
functions of the canonical variables, it thus provides an explicit
example of the quantum adiabatic theorem involving infinitely
degenerate energy levels.

In a previous work \cite{chee}, we obtained a factorization of the
time evolution operator for a charged particle in a slowly rotating
magnetic field with a strong confining potential confining the
particle to be in the plane that is perpendicular to the
instantaneous magnetic field. There, the factorization is valid in
the adiabatic limit only; i.e., no information about the
nonadiabatic factor was obtained due to the complexity of the
problem. The method adopted in the present paper may be applied to
study the nonadiabatic factor in that situation.

\section{Adiabatic perturbations}
It is clear that because of the existence of the factor $M(C_{\bf
R})$ in the time-evolution operator $U(t, 0)$, which gives rise to
the geometric phase phenomenon, one cannot choose a fixed basis of
eigenfuctions of $H_0$, and perform a standard textbook version
perturbative calculation. This is true even if the electric field is
small, because it is the accumulative effects, i.e., $\bf R$ in
$M(C_{\bf R})$, that determines $M(C_{\bf R})$ which is not close to
identity even if $\bf E$ is small.

The factorization of $U(t, 0)$ allows us to do this perturbative
calculation precisely because it identifies a set of
parameter-dependent bases with the help of $M(C_{\bf R})$, then
nonadiabative transition probabilities are completely determined by
$J(C_u)$.

The expression for $J(C_u)$ therefore allows the explicit
calculation of non-adiabatic transition probabilities. Take an
initial eigenstate $|\Psi (n,{\bf R}(0))\rangle=|\Psi (n, {\bf
0})\rangle$ of the Hamiltonian $H_L(0)$ with eigenvalue
$E_n=\hbar\omega(n+1/2)$. The factorization of $U_L$ says that the
time-evolution of $|\Psi (n, {\bf 0})\rangle$ can be seen as the
action of $G(C_{\bf R})D(t)$ on top of $J(C_u)|\Phi(n, {\bf
0})\rangle$. Since $G(C_{\bf R})D(t)$ does not cause transitions,
the transition is caused by the action of $J(C_u)$ on $|\Phi(n, {\bf
0})\rangle$ only. In the adiabatic limit where $|\dot{\bf R}(t)|\sim
u(t)$ is small, one can expand $J(C_u)|\Phi(n, {\bf 0})\rangle$ into
\begin{eqnarray}
J(C_u)|\Phi(n, {\bf 0})\rangle&=&e^{i\gamma(C_u)}(1+\big(\pi
u/\hbar-\pi^\dagger u^\ast/\hbar\big)|\Phi(n, {\bf
0})\rangle)+O(u^2),\nonumber\\&=&e^{i\gamma(C_u)}(1+\big(
uka-ku^\ast a^\dagger\big)|\Phi(n, {\bf 0})\rangle)+O(u^2).\nonumber
\end{eqnarray}
The transition probabilities are
\begin{eqnarray}
|\langle\Phi(n-1, {\bf 0})|J(C_u)|\Phi(n, {\bf 0})\rangle|^2&=&
|\langle\Phi(n-1, {\bf 0})|uka|\Phi(n, {\bf
0})\rangle|^2+O(u^3),\nonumber\\&=&nk^2|u(t)|^2+O(u^3),\nonumber
\end{eqnarray}
\begin{eqnarray}
|\langle\Phi(n+1, {\bf 0})|J(C_u)|\Phi(n, {\bf 0})\rangle|^2&=&
|\langle\Phi(n+1, {\bf 0})|a^\dagger u^\ast/(\hbar k)|\Phi(n, {\bf
0})\rangle|^2+O(u^3),\nonumber\\&=&(n+1)\big(\frac{2qB}{\hbar
c}\big) |u(t)|^2+O(u^3).\nonumber
\end{eqnarray}
All other transition probabilities are zero to the order of
$O(u^2)$. So we see in this case that non-adiabatic transition
probabilities are dependent on the energy levels, and increase with
$n$. This is a result that shows explicitly how the quantum
adiabatic theorem can be realized when an infinitely degenerate
energy level is involved. As usual, the role of the oscillating
$e^{-i\omega s}$ in the expression for $u(t)$ is to make the effect
of $E^{\ast}(s)$ not to accumulate during an adiabatic process with
$t\in[0, T]$, where $T$ is a parameter that $\bf R$ and $\bf E$ may
depend on through ${\bf R}(\frac{t}{T})$ and ${\bf E}(\frac{t}{T})$.
If $E_0/B$ is used as the dimensionless small parameter, say the
electric field has a constant magnitude and that ${\bf E}$ changes
direction slowly compared with $\omega$, then $T$ can be chosen to
be $(B/E)\omega^{-1}$. When $T\gg\omega^{-1}$, one estimates that
$|u(t)|\leq\sim cE/B\omega$, for $t\in[0, T]$. The transition
probabilities are then bounded by $n(\omega^{-1}
c/l_B)^2\cdot(E^2/B^2)$ for $t\in[0, T]$, where $l_B=\sqrt{\hbar
c/(qB)}$ is the magnetic length. For an electron in a $15$ T
magnetic field, if $E=1000$ volts/m, we have
$E/B=1000/(15\times3\times10^8)$, then $n(\omega^{-1}
c/l_B)^2\cdot(E^2/B^2)=n\cdot 1.45\times10^{-5}$, for the duration
of $T=1.71\times10^{-3}$ s. Then for a ground state for example,
excitations can be expected to occur in about $100$ s.

\section{Mixing of the Landau levels beyond the adiabatic limit}
One may also study the mixing of the Landau levels for a general
nonadiabatic time evolution. Since $M(C_{\bf R})D(t)$ in the time
evolution operator does not cause transitions among Landau levels of
$H_0$, the transitions are caused by the action of $J(C_u)$ on
$|\Phi(n, {\bf 0})\rangle$ only. In the most general case, the
expression of $|\langle\Phi(m, {\bf 0})|J(C_u(t))|\Phi(n, {\bf
0})\rangle|^2$ needs to be evaluated in order to determine the
transition probability from an initial state (at $t=0$) that is at
the $n$-th Landau energy level to an $m$-th energy level at time
$t$. We have
\begin{eqnarray}
J(C_u)&=& e^{i\gamma(C_u)}\exp\big(\pi u/\hbar-\pi^\dagger
u^\ast/\hbar\big),\nonumber\\&=&e^{i\gamma(C_u)}\exp\big(uka-u^\ast
k a^\dagger\big).\nonumber
\end{eqnarray}
Using the formula $e^{A+B}=e^{A}e^{B}e^{-\frac{1}{2}[A, B]}$, and
the commutation relation $[a, a^\dagger]=1$, we have
\begin{equation}
J(C_u)=e^{i\gamma(C_u)}e^{-\frac{1}{2}|uk|^2}e^{-u^\ast k
a^\dagger}e^{u k a}.\nonumber
\end{equation}
Therefore, we have the following general expression for the matrix
elements:
\begin{eqnarray}
\langle
m|J(C_u(t))|n\rangle&=&e^{i\gamma(C_u)}e^{-\frac{1}{2}|uk|^2}\big(e^{-u
k a}|m\rangle\big)^{\dagger}\big(e^{u k a}|n\rangle\big),
\end{eqnarray}
which represents mixing of the Landau energy levels for the most
general type of the electric field which is not necessarily small
and which does not have to change slowly.

In particular, this formula implies that for the ground state we
have the following matrix element:
\begin{eqnarray}
\langle
0|J(C_u(t))|0\rangle=e^{i\gamma(C_u)}e^{-\frac{1}{2}|uk|^2},\nonumber
\end{eqnarray}
which implies that if we start with a ground state of $H_0$ at time
0, the probability that it remains to be in a ground state is
$|\langle0|J(C_u(t))|0\rangle|^2=e^{-|uk|^2}=e^{-|u|^2\frac{2qB}{\hbar
c}}$. This implies, for the rotating electric field example as we
discussed earlier where $E=E_0e^{-i\omega t}$ and
$u(t)=-cE_{0}t/(2B)$ at the resonance of $\nu=\omega$, the
probability for the state to remain in a ground state of $H_0$ (not
necessarily the initial ground state because of the existence of the
magnetic translation) is exactly
\begin{equation}
P_{r}(0\rightarrow 0)=\exp[-2(E_0/B)^2(c^2t^2/{l_B}^2)],
\end{equation}
where $l_B=\sqrt{\hbar c/qB}$ is the magnetic length. (Note that for
the electron, $q=-e$, the resonant electric field's angular velocity
is then along the positive $z$ direction, as expected. The
transition therefore can take place extremely fast.)

That the resonance effect exists should be expected from the
physical intuition gained from the classical solution. For the
special case of an electric field along a fixed direction, say a
sinusoidal electric field along the direction of $x_1$, the
transition probabilities can be calculated by choosing the gauge
${\bf A}({\bf x})=(0, Bx,0)$ and the condition $p_2=0$ \cite{budd}
which in effect makes the geometric operator equal to the identity.
For the general situation where the electric field changes
direction, the nonadiabatic factor can be deduced only by first
separating out the geometric operator in the time evolution. This
includes the cases studied in the previous section and in this
section.

\section{Appendix A}
To derive the factorization of $U(t, 0)$, first switch to the
Heisenberg picture. The equations of motion are
\begin{equation}
\dot{\pi}= -i\omega\pi +i \frac{qB}{c}{\dot R}(t), \ \ \ \dot{w}_\mu
= -\frac{qB}{c}\epsilon_{\mu\nu}{\dot R}_{\mu}(t),
\end{equation}
where $\pi =\pi_1+ i\pi_2$, and $R(t)=R_1(t)+ iR_2(t)$. The solution
to the Heisenberg equations can then be expressed as
\begin{equation}
\pi(t)=\pi(0)e^{-i\omega t}+i \frac{qB}{c}e^{-i\omega
t}\int\limits_{0}^{t}e^{i\omega s}\frac{d}{ds} R(s)ds,
\end{equation}
\begin{equation}
w_{\mu}(t)=w_{\mu}(0)- \frac{qB}{c}\epsilon_{\mu\nu}R_{\nu}(t),
\end{equation}
where we assume $R_{\nu}(0)=0$. The homogeneous terms in the
expressions for $\pi(t)$ and $w_{\mu}(t)$ are generated by the usual
dynamical operator $D(t)=\exp(-iH_{0}t/\hbar)$. To produce the extra
terms in the expression for $\pi(t)$, and $w_{\mu}(t)$,
respectively, using an operator $W(t)$, such that $D(t)W(t)$
recovers the whole solution, it suffices for $W(t)$ to satisfy:
\begin{equation}
W^{\dagger}(t)\pi(0)W(t)=\pi(0)+
i\frac{qB}{c}\int\limits_{0}^{t}e^{i\omega s}\frac{d}{ds} R(s)ds,
\nonumber
\end{equation}
\begin{equation}
W^{\dagger}(t)w_{\mu}(0)W(t)=w_{\mu}(0)-
\frac{qB}{c}\epsilon_{\mu\nu}(R_{\nu}(t)-R_{\nu}(0)).\nonumber
\end{equation}
In view of the commutation relations (1), which imply $[\pi,
\pi^{\dagger}]=2\hbar qB/c$, and from the formula
$\exp(-B)A\exp(B)=A+[A,B]$ with the condition that $[A, B]$ commutes
with $A$ and $B$, it is clear that $W(t)$ can be chosen to be the
product of two mutually commuting operators, generated by $(1,
\pi(0), \pi^\dagger(0))$ and $(1, w_{1}(0), w_{2}(0))$ respectively.
Each of theses operators produces a translation for either $\pi(0)$
or $w_{\mu}(0)$ while leaving the other unchanged. Writing $W(t)$ as
$W(t)=J(t)M(t)$, we can make the following choice for $J(t)$ and
$M(t)$,
\begin{equation}
J(t)=T\exp\bigg(i\frac{\pi^{\dagger}(0)}{2\hbar}\int\limits_{0}^{t}e^{i\omega
s}\frac{d}{ds}R(s)ds+
i\frac{\pi(0)}{2\hbar}\int\limits_{0}^{t}e^{-i\omega
s}\frac{d}{ds}R^{*}(s)ds\bigg),
\end{equation}
\begin{equation}
M(t)=P\exp\big(-i{\hbar}^{-1}w_{\mu}(0)R_{\mu}(t)\big),
\end{equation}
where $T\exp$ stands for time-ordered exponential. It's different
from the direct exponential by a numerical phase factor only,
similar to the path-ordered exponential. Therefore, it can be
directly checked that $D(t)J(t)M(t)$ recovers the solutions to the
Heisenberg equations.

To verify that $D(t)J(t)M(t)$ not only recovers the solutions to the
Heisenberg equations for $\pi$ and $w_{\mu}$, but in fact is the
time evolution operator corresponding to $H$, we now verify that it
satisfies the Schr\"{o}dinger equation. Note that $M(t)$ commutes
with both $D(t)$ and $J(t)$, so we have
\begin{equation}
i{\hbar}\frac{\partial}{\partial t}\big(D(t)J(t)M(t)\big)
=i{\hbar}\big[\frac{\partial}{\partial
t}\big(D(t)M(t)\big)\big]J(t)+i{\hbar}M(t)D(t)\frac{\partial}{\partial
t}J(t).  \nonumber
\end{equation}
It is straightforward that
\begin{equation}
i{\hbar}\big[\frac{\partial}{\partial t}\big(D(t)M(t)\big)\big]J(t)
= \big(H_{0}(0)+w_{\mu}\dot{R}_{\mu}(t)\big)D(t)J(t)M(t). \nonumber
\end{equation}
To calculate $i{\hbar}M(t)D(t)\frac{\partial}{\partial t}J(t)$,
first observe that
\begin{equation}
D(t)\pi(0)D^{\dagger}(t)=D^{\dagger}(-t)\pi (0)D(-t)=\pi
(0)e^{i\omega t},
\end{equation}
\begin{equation}
D(t)\pi^{\dagger}(0)D^{\dagger}(t)=\big(D(t)\pi(0)D^{\dagger}(t)\big)^{\dagger}=\pi^{\dagger}(0)e^{-i\omega
t}.
\end{equation}
Therefore
\begin{eqnarray}
i{\hbar}M(t)D(t)\frac{\partial}{\partial t}J(t)&=&\big(-\pi^{\dagger}(0)\dot{R}(t)/2-\pi(0)\dot{R}^{*}(t)/2\big)D(t)J(t)M(t),\\
      &=&-\big(\pi_{1}(0)\dot{R}_{1}(t)+\pi_{2}(0)\dot{R}_{2}(t)\big)D(t)J(t)M(t).
\end{eqnarray}
Combining terms and from the definitions of $\pi_\mu$ and $w_\mu$,
we now have
\begin{eqnarray}
i{\hbar}\frac{\partial}{\partial
t}\big(D(t)J(t)M(t)\big)&=&\big(H_0+\frac{qB}{c}x_1\dot{R}_2-\frac{qB}{c}x_2\dot{R}_1\big)D(t)J(t)M(t),\\
                        &=&H\big(D(t)J(t)M(t)\big).
\end{eqnarray}
Therefore, we conclude that the time evolution operator
corresponding to $H$ is
\begin{equation}
U(t, 0)=D(t)J(t)M(t)=M(t)D(t)J(t).
\end{equation}

\end{document}